\documentclass[aps,pra,twoside,twocolumn,10pt]{revtex4-2}
\usepackage[colorlinks=true, citecolor=blue, urlcolor=blue ]{hyperref}
\usepackage{epsfig,newlfont,amssymb,amsfonts,amsmath,bm,subfigure,palatino,mathtools,amsthm,braket,soul,enumitem,color,graphics,graphicx,times,physics}
\usepackage[normalem]{ulem}
\graphicspath{{Plots/}{Dynamics/}}

\begin{document}
\title{
Recognizing critical lines via entanglement in non-Hermitian systems
}

\author{Keshav Das Agarwal$^1$, Tanoy Kanti Konar$^1$, Leela Ganesh Chandra Lakkaraju$^{1,2,3}$, Aditi Sen (De)$^1$}

\affiliation{$^1$ Harish-Chandra Research Institute, A CI of Homi Bhabha National Institute,  Chhatnag Road, Jhunsi, Allahabad - 211019, India}
\affiliation{$^2$ Pitaevskii BEC Center, CNR-INO and Dipartimento di Fisica, Universit\`a di Trento, Via Sommarive 14, Trento, I-38123, Italy }
\affiliation{$^3$ INFN-TIFPA, Trento Institute for Fundamental Physics and Applications, Trento, Italy} 

\begin{abstract}

The non-Hermitian model exhibits counterintuitive phenomena that are not observed in the Hermitian counterparts. To probe the competition between non-Hermitian and Hermitian interacting components of the Hamiltonian, we focus on a system containing non-Hermitian $XY$ spin chain and Hermitian Kaplan-Shekhtman-Entin-Aharony (KSEA) interactions along with the transverse magnetic field.  We show that the non-Hermitian model can be an effective  Hamiltonian of a Hermitian $XX$ spin-$\frac{1}{2}$ with KSEA interaction and a local magnetic field that interacts with local and nonlocal reservoirs. The analytical expression of the energy spectrum divides the system parameters into two regimes: in one region, the strength of Hermitian KSEA interactions dominates over the imaginary non-Hermiticity parameter, while in the other, the opposite is true. In the former situation, we demonstrate that the nearest-neighbor entanglement and its derivative can identify quantum critical lines with the variation of the magnetic field. In this domain, we determine a surface where the entanglement vanishes, similar to the factorization surface, known in the Hermitian case. On the other hand, when non-Hermiticity parameters dominate, we report the exceptional and critical points where the energy gap vanishes and illustrate that bipartite entanglement is capable of detecting these transitions as well. Going beyond this scenario, when the ground state evolves after a sudden quench with the transverse magnetic field, both the rate function and the fluctuation of bipartite entanglement quantified via its second moment can detect critical lines generated without quenching dynamics. 



\end{abstract}

\maketitle


\section{Introduction}

Non-Hermitian quantum systems provide a fresh perspective on traditionally established concepts in the Hermitian domain such as quantum phase transitions, topological phases, and the role of symmetry in quantum mechanics \cite{gong_prx_2018,kawabata_prx_2019,chen_njp_2019}. For almost two decades or so, investigations of such systems have generated lots of interest since effective non-Hermitian systems can be obtained by dilating the system in a larger Hermitian system \cite{naimark_dilation_2008} or by introducing skew Hermitian part in the dynamics of a state which include gain and loss of energy or particles \cite{brody_prl_2012,liu_prap_2020}. Moreover, when the system is in contact with a noisy environment, the evolution of the system, in the weak coupling limit, is well described by the Gorini–Kossakowski–Lindblad-Sudarshan  (GKLS) \cite{open_quan_book, lidar_2020_lecture} master equation involving Lindblad operators which leads to non-Hermitian Hamiltonian \cite{ohlsson_pra_2021,khandelwal_prx_2021,chen_prl_2022,nakanishi_pra_2022} and quantum jump operators realized by the continuous measurement performed on the system by the environment \cite{minganti_pra_2020,fleckenstein_prr_2022} . 
Further, the spectacular experimental  developments in several physical platforms like photonics \cite{ozdemir_np_2019}, and superconducting systems \cite{chen_prl_2021} promise to realize such non-Hermitian systems in a controlled manner.
As a result, various suggestions for quantum technologies, such as quantum thermal machines \cite{khandelwal_prx_2021, konar2022quantum}, and quantum sensors \cite{ep_sensing_1, chen_njp_2019, ep_sensing_3,ep_sensing_4, ep_sensing_5, chen_nature_ep_sensing_2017, naghiloo_nature_ep_sensing_2019}, entanglement purification~\cite{gopalakrishnan_prl_2021}  that exploit non-Hermiticity in systems, have been developed in recent years.

In this work, we concentrate on the non-Hermitian Hamiltonian, which is an effective Hamiltonian in the framework of open quantum systems through the quantum trajectory method without quantum jump operators \cite{Ueda_review,minganti_pra_2020,fleckenstein_prr_2022, abbasi_prl_2022}. Such systems possess exceptional points (EPs) where the corresponding Hamiltonian becomes defective due to the coalescence of eigenvectors \cite{bender_prl_1998,brody_iop_2014,Franco_nori_lep_hep_2019} (see Refs. \cite{Franco_nori_lep_hep_2019, chen_prl_2022} for  Liouvillian EPs). It was shown that the EPs of the non-Hermitian Hamiltonian with parity-time reversal symmetry (\(\mathcal{PT}\)-symmetry) can divide the parameter space into  two phases :  unbroken phase with  real eigenspectrum which can be mapped to a  Hermitian Hamiltonian \cite{mostafazadeh_1,bender_prl_2002},  and the broken region having complex eigenvalues. The presence of non-Hermiticity can often result in counter-intuitive behaviors such as localization-delocalization phenomena  \cite{yang_prb_2017,hamazaki_prl_2019}, dynamical quantum phase transitions (DQPT) \cite{heyl_prl_2013, heyl_prl_2015,Heyl_review_2018, jafri_sr_2019_9,jafri_prb_2019_10, stav_prb_2020,jafri_prb_2020_5, jafri_iop_2020_7,modak_prb_2021, jafri_pra_2021_6,jafri_prb_2021_4,haldar_prx_2021, nandi_prl_2022, jafri_pra_2022_3, jafri_prb_2022_2, jafri_prb_2022_1} in non-Hermitian lattices \cite{zhou_pra_2018}, gapless phases in $p$-wave superconductors exhibiting interesting features in dynamics \cite{zhou_pra_2018, yang_pra_2020,mondal_prb_2022, mondal_arxiv_2022},  multiparty entanglement in the Lipkin-Meshkov-Glick model \cite{lee_prl_2014}, quantum phase transition in the dissipative Ising chain \cite{guo_pra_2022, yang_njp_2022}, weak ergodicity breaking in many-body system \cite{chen_arxiv_2023} and mode entanglement between fermions differentiating between nontrivial and trivial phases of topological materials \cite{li_prb_2018, yamamoto_prl_2019,loic_scipost_2019, xiao_prb_2022, turkeshi_prb_2023} and differentiating volume and area law of entanglement in free-fermionic system ~\cite{kawabata_prx_2023,gal_scipost_2023}.
 
\begin{figure}
     \centering
    \includegraphics[width=0.9\linewidth]{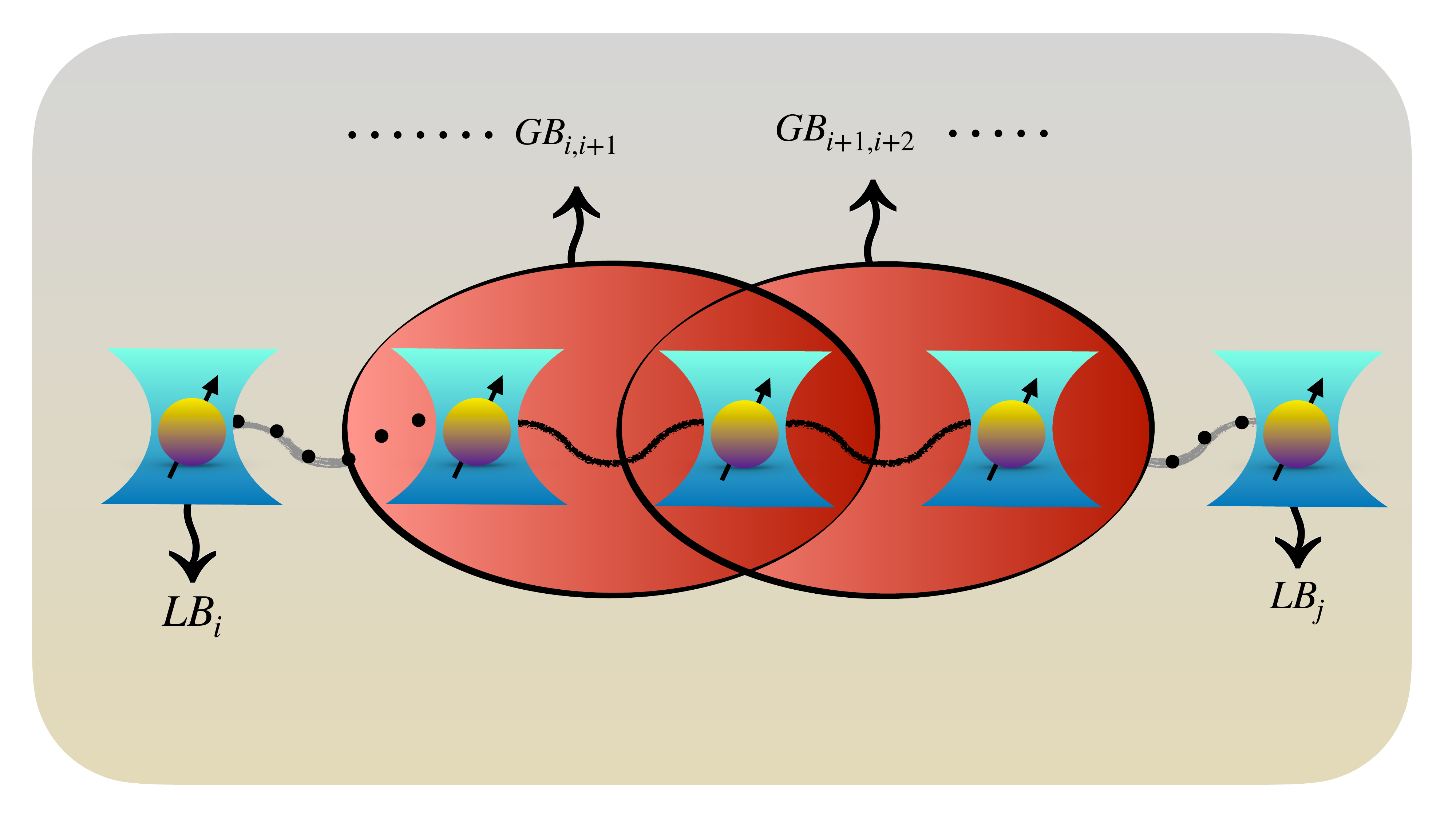}
    \caption{ Schematic representation of the effective non-Hermitian Hamiltonian. When the $XX$ model with KSEA interactions between neighboring sites and transverse magnetic fields are in contact with local (LB) and non-local baths (GB), the dynamics can be governed by the effective non-Hermitian model, studied in this work.  It leads to a possible method to realize this Hamiltonian in laboratories. }
    \label{fig:sch}
\end{figure}

 In recent years, the broken-to-unbroken transitions of non-Hermitian quantum spin models were studied in which the non-Hermiticity is inserted either in the interaction part by making imaginary coupling constant or by taking the imaginary strength of the magnetic field in the Hamiltonian \cite{Song_RT_symm,zhang_pra_2014,zhang_pra_2016}. 
In this paper, we focus on an interacting quantum spin model, which can activate a competition between Hermitian and non-Hermitian interaction terms. In particular, we introduce a rotational and time-reversal symmetric system consisting of non-Hermitian interaction parts in the \(xy\)-plane having an imaginary non-Hermiticity constant, representing \(iXY\) model and a Hermitian Kaplan-Shekhtman-Entin-Aharony interaction (KSEA) \cite{kaplan_jpb_1983,shekhtman_prl_1992,zheludev_prl_1998} in presence of a magnetic field in the \(z\)-direction.  First, we demonstrate that such a non-Hermitian Hamiltonian is an effective description of a spin chain with \(XX\)  and KSEA interactions having a transverse magnetic field attached to a local and non-local bath, which can be achieved by manipulating the baths \cite{metelmann_prx_2015, Fazio_reservoir_engineering_2018} (see Fig. \ref{fig:sch}).  
Moreover, the Hamiltonian can be mapped to a quadratic free-fermionic model using the Jordan-Wigner transformation \cite{LSM_main, barouch_pra_1970, barouch_pra_1971, fu_epjb_2020} and hence the behavior of several physical quantities including energy, classical correlators, and bipartite entanglement can be investigated both for finite-size systems consisting of a large number of sites and in the thermodynamic limit. In an experimental set-up, the two-qubit reduced density matrix can be obtained via tomography, by measuring local observables like magnetization and two-site classical correlators. Therefore, the entanglement between nearest-neighbor sites can  be computed via post-processing. Various works in the last decade quantify correlators and entanglement in experimental set-ups~\cite{Zhang2011, Foroozani2016, Struchalin2016, Jeng2019, MllerRigat2022}. In terms of numerics,  while computation of (complex) eigenvalues of the entire spectra is a daunting task, we can avoid such computation by considering the nearest-neighbor entanglement corresponding to the minimum eigenenergy.

We indeed observe a trade-off relation between imaginary non-Hermiticity parameters and the interaction strength of KSEA. Specifically, we exhibit that the entire analysis can be divided into two distinct regions :  in one region, the strength of the KSEA interaction is slightly higher than the non-Hermiticity parameter present in the \(iXY\) part; in the other domain,  the non-Hermiticity parameter rules over the KSEA interaction strength, thereby increasing the non-Hermiticity component in the system. In the former case, an entire unbroken region emerges where the system mimics the properties of a Hermitian Hamiltonian, known as the \(\mathcal{RT}\)-symmetry-protected region. In particular, the model in this case contains two striking features -- quantum critical point obtained by tuning the strength of the magnetic field where the energy gap vanishes and the factorization surface where the entanglement vanishes. We report that the entanglement pattern, especially the derivative of the nearest-neighbor entanglement, can capture both critical points and factorization surfaces. Moreover, we show that by gauge transformation in the fermionic basis,  the non-Hermitian model having both \(XY\) and KSEA interactions can be transformed into the Hermitian \(XY\) model having a factorization surface known in the literature \cite{giampaolo_prl_2008, giampaolo_prb_2009} which matches with the factorization surface found in the non-Hermitian counterparts. 
In the case where non-Hermiticity dominates, we analytically determine that the system possesses an exceptional point, separating broken phase from the unbroken one and also quantum critical points having vanishing gap from the expression of the energy spectrum.  By carefully defining the right eigenvector having the lowest imaginary eigenvalue, we demonstrate that the derivative of nearest-neighbor entanglement of the ground state can show non-analyticity around the exceptional point as well as at the critical point with respect to the magnetic field. It establishes that bipartite entanglement can capture aspects that are solely present in the non-Hermitian model as well as critical points connected to gap-closing.


As non-Hermitian systems typically arise in non-equilibrium situations, it is interesting to probe the system with a sudden quench \cite{agarwal2022detecting}. Initially, the system is prepared in the ground state of the unbroken region and is then quenched in various parameter regimes, especially across the critical points based on gap-closing. Specifically, we illustrate that when the quenching is across the critical point, i.e., the initial magnetic field is much stronger than the strength of the final magnetic field and vice-versa,  the conventional dynamical quantity like the rate function based on the overlap of the initial and the evolved states shows non-analyticity with time, thereby detecting the quantum critical point. However, there are situations where both the initial and quenched magnetic fields are weak, the rate function can be non-analytic, giving a false signal of crossing the quantum critical point.  
In addition to the rate function, we examine the dynamics of nearest-neighbor entanglement, quantified by its second moment, and determine that it is capable of identifying critical points in the system.

The organization of the paper is as follows. In Sec. \ref{sec:effect_hamil}, we introduce the model having $iXY$ and KSEA interactions along with the magnetic field and provide the Hermitian model which leads to its effective description. In Sec. \ref{sec:excep_fac}, we analytically derive the exceptional point and establish its connection with the factorization surface of the corresponding Hermitian model. We demonstrate the capability of nearest-neighbor entanglement in detecting critical points present in this model in Sec. \ref{sec:static_entang}. In Sec. \ref{sec:dynamics},  by performing sudden quench of the ground state in different parameter regimes, the dynamical states are studied with the help of Loschmidt echo, and the corresponding rate function as well as via nearest-neighbor entanglement in Secs. \ref{sec:losh_echo} and  \ref{sec:dyn_ent} respectively. Finally, we conclude in Sec. \ref{sec:conclu}.

\section{Effective non-Hermitian Hamiltonian}
\label{sec:effect_hamil}

In general, non-Hermiticity in a quantum system can be attributed to two different scenarios : $(1)$ continuous measurement on the system, and $(2)$ the system undergoing dissipative dynamics due to the interaction with the environment which can effectively be described by a non-Hermitian Hamiltonian. These kinds of non-Hermitian Hamiltonian are useful as they possess real eigenvalues, known as pseudo-Hermitian models, having some symmetry. Such quintessential examples of anti-unitary symmetries which bestow a Hamiltonian with real eigenvalues are \(\mathcal{PT}\) and \(\mathcal{RT}\)-symmetries, where \(\mathcal{P}\), \(\mathcal{R}\) and $\mathcal{T}$ stand for parity, rotation and time reversal operation respectively. A region containing a set of parameters in the system having imaginary energy spectrum is called broken while tuning a set of parameters leads to a real spectrum in a non-Hermitian model, known as the unbroken phase.  The transition from broken to the unbroken region by varying system parameters occurs around a transition point, referred to as an exceptional point, which plays a crucial role in quantum metrology, thermal engines, and many more \cite{chen_nature_ep_sensing_2017,chen_njp_2019,khandelwal_prx_2021}.

In this paper, we focus on the \(\mathcal{RT}\)-symmetric Hamiltonian where \(\mathcal{R}\equiv \exp[-i\frac{\pi}{4}\sum_{j=1}^N\sigma_j^z]\), a \(\frac{\pi}{2}\) rotation around the \(z\)-axis and the time-reversal operation, \(\mathcal{T}i\mathcal{T}^{-1}=-i\). Although the Hamiltonian satisfies the relation, \([H,\mathcal{RT}]=0\), due to the anti-unitary properties of the symmetry operations in the broken region, eigenstates of the Hamiltonian do not have \(\mathcal{RT}\) symmetry. Let us describe how the non-Hermitian Hamiltonian containing imaginary interaction strength described below can be written as an effective description of a system evolving and interacting with an engineered reservoir. Notice that an effective Hamiltonian undergoing continuous monitoring or measurements can be shown to represent a non-Hermitian system in the presence of imaginary magnetic field having $\mathcal{P}$ and $\mathcal{T}$ symmetry \cite{lee_prx_2014, turkeshi_prb_2023, turkeshi_prb_2023b}. 

We consider a one-dimensional spin model of $N$ spin-$\frac{1}{2}$ particles with nearest-neighbor interactions, governed by the Hamiltonian described as
\begin{eqnarray}
    \nonumber H_{S}&=&J\sum_{j}(\sigma_j^x\sigma_{j+1}^x+\sigma_j^y\sigma_{j+1}^y)+\frac{h^\prime}
{2}\sigma_j^z\\&&+\frac{K^\prime}{4}(\sigma_j^x\sigma_{j+1}^y+\sigma_j^y\sigma_{j+1}^x)
    \label{eq:ksea_xy}
\end{eqnarray}
where $\sigma^\alpha (\alpha=x,y,z)$ denotes the Pauli matrix with the periodic boundary being imposed, $\sigma_{N+1}=\sigma_1$, the first two interacting terms represent the \(XX\) interaction, $\frac{h^\prime}{J} = h$ is the strength of the transverse magnetic field, and $\frac{K^\prime}{J}=K$ is the coupling parameter of the symmetric helical interaction, called Kaplan-Shekhtman-Entin-Aharony  interaction \cite{kaplan_jpb_1983,shekhtman_prl_1992,zheludev_prl_1998} which is found in solid-state materials such as \(\text{Ba}_2\text{CuGe}_2\text{O}_7\) and \(\text{Yb}_4\text{As}_3\) \cite{chovan_prb_2013}.

Reservoir engineering has been instrumental in many phenomena, such as nonreciprocal photon transmission \cite{metelmann_prx_2015} and persistent currents in the system \cite{Fazio_reservoir_engineering_2018}. Suppose every individual spin is attached to a local bath while each nearest-neighbor spin pair is attached to a common non-local bath (see schematic representation of the set up in Fig. \ref{fig:sch}). In this scenario, the evolution of the system interacting with an engineered reservoir is governed by the GKLS master equation \cite{open_quan_book}, given by
\begin{equation}
    \frac{d\rho}{dt}=-i[H_{S},\rho]+\kappa \sum_{j}\mathcal{L}[\sigma_j^-](\rho)+\sum_{j}\mathcal{L}[g_j(\{\sigma\})](\rho),
    \label{eq:master_eqn}
\end{equation}
where \(\mathcal{L}[A]\) are a set of operators that effectively describe the effects of the environment on the system, known as the Lindblad operators. Note that the second term, \(\mathcal{L}[\sigma_j^-]\) represents the local dissipation of strength \(\kappa\) and the third term is present due to non-local dissipation between the sites represented as \(\mathcal{L}[g_j(\{\sigma\})]\), where \(g_j(\{\sigma\})\) is given as
\begin{equation}
    g_j(\{\sigma\})=p \sigma_j^-+q \sigma_j^++r\sigma_{j+1}^-+s\sigma_{j+1}^+ ,
    \label{eq:lidblad}
\end{equation}
with \(p,q,r \text{ and }s\) denoting the suitable coupling parameters with the correlated environment, which can, in general, be complex.
\cite{Fazio_reservoir_engineering_2018}. The action of such superoperators is given mathematically as \(\mathcal{L}[A](\rho)=A\rho A^\dagger-\frac{1}{2}\{A^\dagger A,\rho\}\) where \(A\rho A^\dagger\) is a jump operation. In order to arrive at a \(\mathcal{RT}\)-symmetric Hamiltonian, we choose \(p,s=0\), \(q=-\sqrt{\gamma}\), and \(r=\sqrt{\gamma}\) and Eq. (\ref{eq:master_eqn}) reduces to
 \begin{eqnarray}
     \nonumber\frac{d\rho}{dt}&=&-i(H_{\text{eff}}\rho-\rho H_{\text{eff}}^\dagger)+\sum_{j}\mathcal{L}[A_j](\rho) ,
 \end{eqnarray}
 where \(H_{\text{eff}}=(H_S-\frac{i}{2}\sum_j \kappa\sigma_j^+\sigma_j^-+ g_j(\{\sigma\})^\dagger g_j(\{\sigma\}))\) is the effective Hamiltonian which is non-Hermitian. In the semi-classical limit, neglecting the jump operation (no-click limit in the continuous measurement setting), the evolution of the system can be shown to be governed by the non-Hermitian Hamiltonian. Hence, the effective Hamiltonian is given by

\begin{eqnarray}
    \nonumber H_{\text{eff}}=&&H_{S}-\frac{i}{2}\sum_{j=1}^{N}g_j(\{\sigma\})^\dagger g_j(\{\sigma\})\\
    \nonumber =&&H_{S}-\frac{i\gamma}{2}\sum_{j=1}^{N}(\sigma_j^+-\sigma_{j+1}^-)(\sigma_j^--\sigma_{j+1}^+)\\
    =&& \sum_{j=1}^{N} h\sigma_j^z +\frac{i\gamma}{2}+ \frac{1}{4}\Big ((1+i\gamma)\sigma_j^x\sigma_{j+1}^x\\
    \nonumber&&+(1-i\gamma)\sigma_j^y\sigma_{j+1}^y\Big)+\frac{K}{4}(\sigma_j^x\sigma_{j+1}^y+\sigma_j^y\sigma_{j+1}^x).
\end{eqnarray}
Notice that there are two imaginary terms present, one in the coupling and another one is a constant term, in the Hamiltonian, thereby making it non-Hermitian. By modifying local disspative environment strength $\kappa$ in Eq. (\ref{eq:master_eqn}), the imaginary part of the magnetic field can be made vanishing, so that the effective Hamiltonian only contains the imaginary part in the coupling, leading to \cite{zhang_arxiv_2023} 
\begin{eqnarray}
    \nonumber H^{\text{iKSEA}}&=&\sum_{j}\frac{(1+i\gamma)}{4}\sigma_j^x\sigma_{j+1}^x+\frac{(1-i\gamma)}{4}\sigma_j^y\sigma_{j+1}^y\\&&+\frac{K}{4}(\sigma_j^x\sigma_{j+1}^y+\sigma_j^y\sigma_{j+1}^x)+\frac{h}{2}\sigma_j^z,
\end{eqnarray}
with $\gamma$ being the non-Hermiticity parameter. We will illustrate that the above non-Hermitian system possesses some rich properties which we will reveal with and without quenching dynamics.

\begin{figure}
     \centering
    \includegraphics[width=\linewidth]{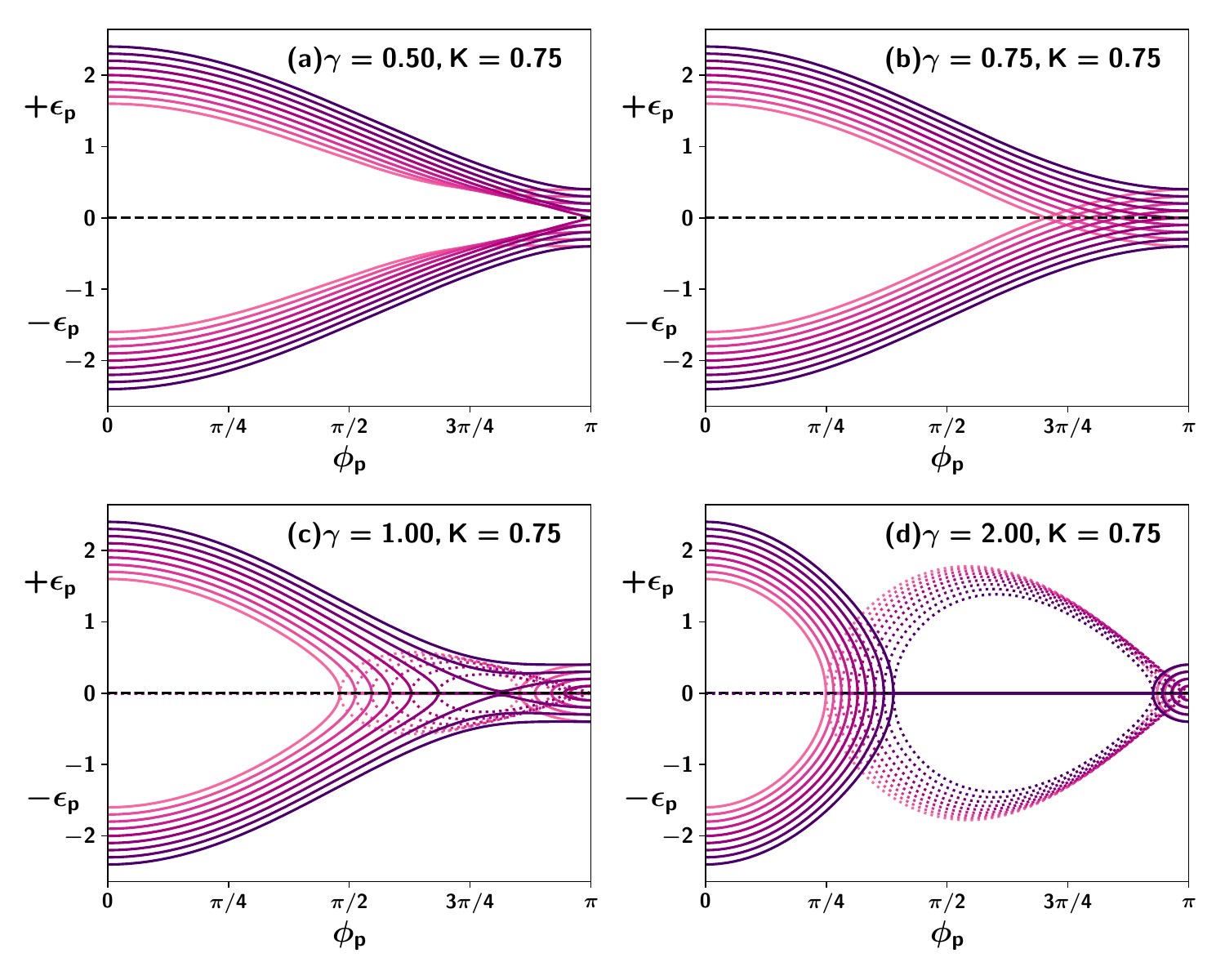}
    \caption{{\bf Energy, \(\pm \epsilon_p\) (ordinate) of the iKSEA model against \(\phi_p\). }  (a) and (b). \(\gamma \leq K\) , i.e., Hermitian KSEA interactions dominates over non-Hermiticity parameter while in (c) and (d), the opposite picture is considered with \(\gamma >k\). Different lines represent different strength of the magnetic field, \(0.6\) (lightest)\(\leq h\leq 1.4\) (darkest). Solid and dotted lines indicate real and imaginary energies respectively. All  axes are dimensionless. 
    }
    \label{fig:energy_iksea}
\end{figure}

\subsection*{Exceptional point and connection with factorization point}
\label{sec:excep_fac}

In order to describe the distinctive features of \(H^{\text{iKSEA}}\), we need to identify the exceptional point which can be obtained by evaluating the dispersion relation for the non-Hermitian Hamiltonian \cite{Song_RT_symm}. 
By applying the Jordan-Wigner transformation (JW), the model reduces to a free-fermionic Hamiltonian given by
\begin{eqnarray}
    \nonumber H^{\text{iKSEA}}_{JW}&=&\frac{1}{2}\sum_j (c_j^\dagger c_{j+1}+c_{j+1}^\dagger c_{j})+ i(\gamma-K)c_j^\dagger c_{j+1}^\dagger\\&&+i(\gamma+K)c_{j+1} c_j+h(2c_j^\dagger c_{j}-1),
    \label{eq:iksea_jw}
\end{eqnarray} 
where $c_j^ \dagger$ and $c_j$ represent fermionic creation and annihilation operators, following $\{c_i, c_j^\dagger \} = \delta_{ij}$. The above model is also known as the Kitaev model with imbalanced pairing, with exotic topological properties \cite{kitaev_iop_2001, li_prb_2018}. Let us perform the second step involving Fourier transformation which transforms the fermionic operators into their conjugate momentum as
\begin{eqnarray}
    \nonumber c_j&=&\frac{1}{\sqrt{N}}\sum_{p=-N/2}^{N/2}\exp(-i\frac{2\pi jp}{N})a_p, \\  \text{ and }\,\,  c_j^\dagger &=&\frac{1}{\sqrt{N}}\sum_{p=-N/2}^{N/2}\exp(i\frac{2\pi jp}{N})a_p^\dagger.
\end{eqnarray}
Due to the periodic boundary condition, the system possesses translational invariance, such that the momentum is a good quantum number, which allows the Hamiltonian to be decomposed into the individual momentum sectors, described by \(H^{\text{iKSEA}}_{JW}=\oplus H_p^{\text{iKSEA}}\). Therefore, Eq. (\ref{eq:iksea_jw}) reduces to
\begin{eqnarray}
    \nonumber \bar{H}^{\text{iKSEA}}_p &=&  \sum_{p>0} (h+\cos\phi_p)(a_p^\dagger a_p+a_{-p}^\dagger a_{-p})\nonumber\\&&+\sin\phi_p[(\gamma -K) a_p^\dagger a_{-p}^\dagger+(\gamma +K) a_pa_{-p}]-h, \nonumber
    \label{pksea_ixy}
\end{eqnarray}
where \(\phi_p=(2p-1)\pi /N\) and \(p\in {1,\ldots, N/2}\) as we choose anti-periodic boundary condition, \(c_{N+1}=-c_1\) \cite{santoro_ising_beginners_2020}. In the thermodynamic limit, i.e., \(N\to\infty\), the momentum becomes continuous, and we have \(\phi_p\in (0,\pi)\). Writing $H^{\text{iKSEA}}$ in the basis, $\{|0\rangle, a^\dagger_p a^\dagger_{-p}|0\rangle, a^\dagger_p|0\rangle,a^\dagger_{-p}|0\rangle\}$, we have
\small{
\begin{align}
\bar{H}^{\text{iKSEA}}_p =\left[\begin{array}{cccc}
-h & -(\gamma +K) \sin \phi_p & 0 & 0 \\
(\gamma -K) \sin \phi_p  & 2\cos \phi_p+h & 0 & 0 \\
0 & 0 & \cos \phi_p & 0 \\
0 & 0 & 0 & \cos \phi_p
\end{array}\right].
\label{eq:ixy_ksea_p}
\end{align}}
The Hamiltonian in Eq. (\ref{eq:ixy_ksea_p}) can be written as $\bar{H}_p^{\text{iKSEA}} = \hat{A}_p^\dagger H_p^{\text{iKSEA}} \hat{A}_p,$ where \(\hat{A}_p\) is the column vector, \((a_{-p}^\dagger, a_p)^T\), also known as the Nambu spinor \cite{santoro_ising_beginners_2020}, and 
\begin{equation}
    H_p^{\text{iKSEA}}=\left[\begin{array}{cc}
-h-\cos \phi_p & -(\gamma+K) \sin \phi_p \\
(\gamma-K) \sin \phi_p & \cos \phi_p+h
\end{array}\right].
\label{eq: ksea_P}
\end{equation}
The eigenvalues of this model are $E_p^{\text{iKSEA}} = \pm\epsilon_p$, where 
\begin{equation}
    \epsilon_p=\sqrt{(h+\cos\phi_p)^2-(\gamma^2-K^2)\sin^2\phi_p}.
    \label{eq: dispersion_relation}
\end{equation}
Depending on the values of \(\gamma\) and \(K\), two separate regions emerge (see Fig. \ref{fig:energy_iksea}).

\textbf{Region-I: $\gamma> K$.} In this region, we first notice that the second term in Eq. (\ref{eq: dispersion_relation}) is positive and hence when it dominates over the first term, the system can have imaginary eigenspectrum. To determine the exceptional point, by calculating \(\pdv{\epsilon_p}{\phi_p}=0\) and \(\epsilon_p=0\) from Eq. (\ref{eq: dispersion_relation}), we obtain the exceptional point, \(h_{\text{EP}}=\sqrt{1+\gamma^2-K^2}\). Note that the exceptional point for the $iXY$ model with $K=0$ is $\sqrt{1+\gamma^2}$, which is obtained in the previous study \cite{Song_RT_symm}. On the other hand, if we consider the Hermitian KSEA model \cite{fu_epjb_2020} where we replace $i\gamma$ by $\gamma$, the Hamiltonian reads as
\begin{eqnarray}
    \nonumber H^{KSEA}&=&\sum_{j}\frac{(1+\gamma)}{4}\sigma_j^x\sigma_{j+1}^x+\frac{(1-\gamma)}{4}\sigma_j^y\sigma_{j+1}^y\\&&+\frac{K}{4}(\sigma_j^x\sigma_{j+1}^y+\sigma_j^y\sigma_{j+1}^x)+\frac{h}{2}\sigma_j^z.
\end{eqnarray}
By applying a similar procedure as described above, the energy eigenvalues can be obtained as $E_p^{\text{KSEA}} = \pm\sqrt{(h+\cos\phi_p)^2+(\gamma^2+K^2)\sin^2\phi_p}$ \cite{barouch_pra_1970}. It can be shown that the ground state of the Hamiltonian is a fully product state that has vanishing bipartite and multipartite entanglement where \(h_{f}^{\text{KSEA}}=\sqrt{1-\gamma^2-K^2}\) (see Appendix. \ref{sec: factorization_KSEA}) known as the factorization surface. Hence, the surface of EP is connected to the factorization surface if the non-Hermiticity parameter is changed to imaginary to make the Hamiltonian Hermitian \cite{ganesh_aditi_factorization_surface}.

\textbf{Region-II: $\gamma\le K$.} In this situation, the second term is always non-positive and, therefore, there exists no momentum block where the model possesses complex eigenvalues. This sector is protected by \(\mathcal{RT}\)-symmetry without having a broken-to-unbroken transition. 

However, the model has different characteristics with the variation of the magnetic field strength. Specifically, analyzing Eq. (\ref{eq: dispersion_relation}), we observe that the extremum momenta (i.e., \(\phi_p = \pi\)) indicate a quantum critical point at $h_c = 1$. This is the typical second-order quantum phase transition (QPT) \cite{sachdev_2011} that is observed in the transverse $XY$ model in the Hermitian domain. 
As reported for the Hermitian $XY$ model, we will now show that bipartite entanglement is capable to identify criticalities in the non-Hermitian model as well. It is important to stress here that the computation of bipartite entanglement for this model requires more careful treatment than its Hermitian counterpart, as will be illustrated in the succeeding section.  

\section{Identifying criticalities via bipartite entanglement}
\label{sec:static_entang}

The variation of classical correlations between two sites with the driving parameter of the Hamiltonian is historically used to signal the transition between phases at zero-temperature of the Hermitian systems. 
Around two decades ago, it was shown that nearest-neighbor quantum correlations, specifically entanglement can also determine the quantum critical point. In this respect, our contribution here is to establish the power of bipartite entanglement in detecting quantum critical points in non-Hermitian systems. 
In order to define the bipartite density matrix in the case of a non-Hermitian system, we have to address a certain subtlety. As the Hamiltonian is non-Hermitian, there are right and left eigenvectors corresponding to an eigenvalue. Thus, a density matrix can be defined in many ways including the left and right vectors. However, in this work, we define $\rho = \frac{|R\rangle\langle R|}{\text{Tr} \rho}$ \cite{lee_prx_2014, turkeshi_prb_2023, turkeshi_prb_2023b}, where $|R\rangle$ is the right eigenvector of the ground state. In case of calculations of experimentally feasible observables, such a choice is valid since the density matrix becomes Hermitian. A detailed description of the ground state assumption is explained in Appendix~\ref{sec:ground_state_non_herm}.
 
From Eq. (\ref{eq: ksea_P}), we rewrite each block as
\begin{equation}
    H^{\text{iKSEA}}_p=\hat{A}_p^\dagger V_p^{-1}V_p H^{\text{iKSEA}}_p V_p^{-1}V_p \hat{A_p},
\end{equation}
where \(V_pH^{\text{iKSEA}}_p(V_p)^{-1}=\text{diag}(+\epsilon_p,-\epsilon_p)\). We define a new operator, \(\hat{\eta}=V_p\hat{A_p}\), which forms the non-Hermitian Bogoliubov basis. The matrix \(V_p\) can be established by the right eigenvectors and the non-Hermitian Bogoliubov transformation can be represented as \cite{lee_prx_2014} 
\begin{eqnarray}
    \nonumber \eta_p&=&-u_pa_{-p}^\dagger-v_p^1 a_{p} \\
    \text{ and } \quad 
    \bar{\eta}_p&=&-v_p^2 a_{p}^\dagger+u_pa_{-p}, 
\end{eqnarray}
where
\begin{equation*}
    u_p=\frac{h+\cos\phi_p-\epsilon_p}{\sqrt{M}},\quad v_p^1=\frac{(\gamma+K)\sin\phi_p}{\sqrt{M}},
\end{equation*}
and
\begin{equation}
    v_p^2=\frac{(\gamma-K)\sin\phi_p}{\sqrt{M}},
\end{equation}
with \(M=-{(h+\cos\phi_p-\epsilon_p)^2}+(\gamma^2-K^2)\sin^2\phi_p\) and \(\{\eta_p,\bar{\eta}_{p'}\}=\delta_{pp'}\). 
\begin{figure}
     \centering
    \includegraphics[width=\linewidth]{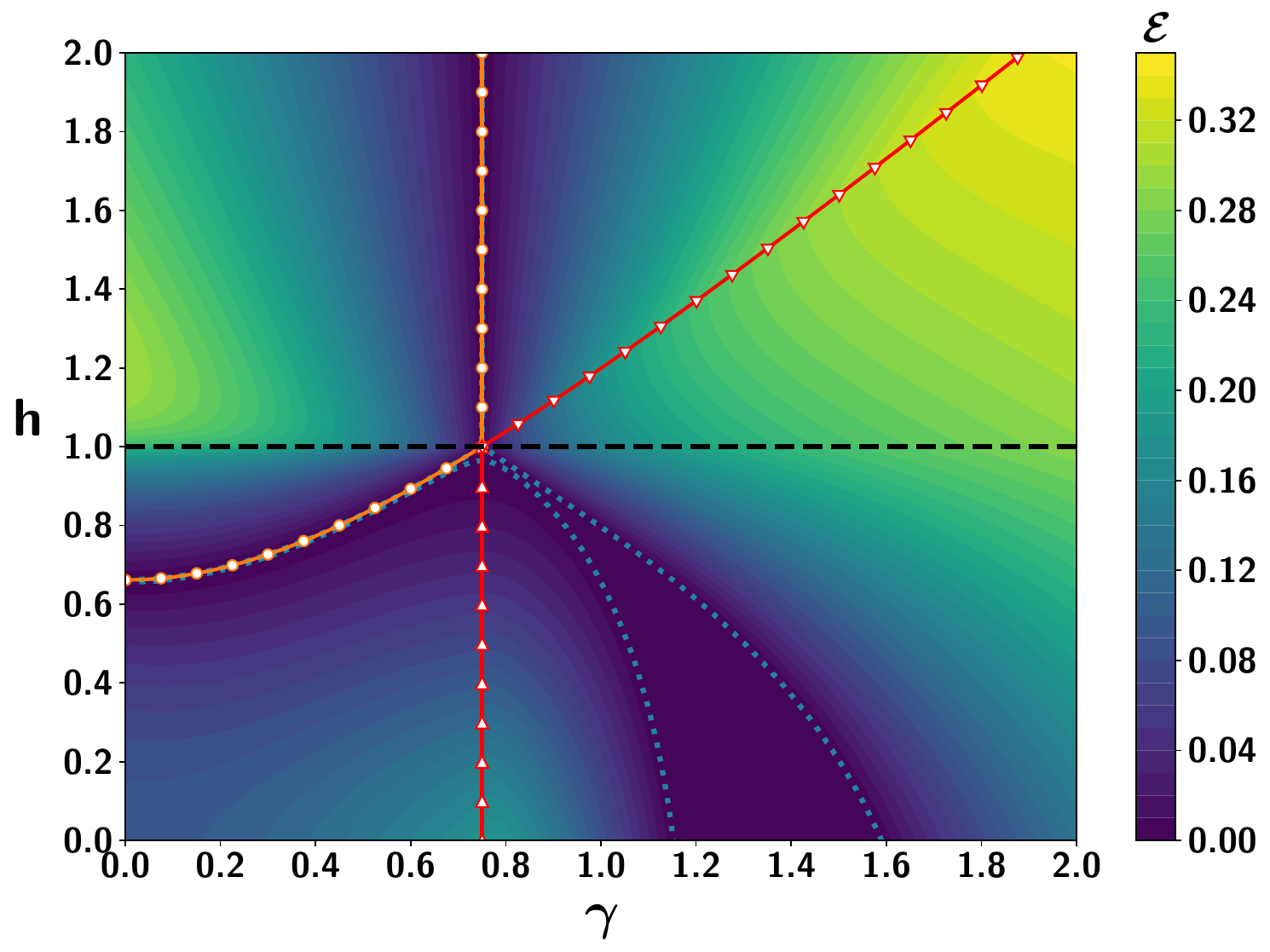}
    \caption{Contour plot of nearest-neighbor entanglement  against external magnetic field, \(h\) (vertical axis) and non-Hermiticity parameter, \(\gamma\) (horizontal axis) of the spin chain with both non-Hermitian XY and  Hermitian KSEA interactions, given in Eq. (\ref{eq:ksea_xy}).  Here we set \(K=0.75\) and \(N=5000\). Lines with triangles and circles represent the exceptional points and factorization surface respectively, while the dashed line represents the critical line. All axes are dimensionless. 
    }
    \label{fig:ent2}
\end{figure}

If \(\ket{G}\) is the ground state of the Hamiltonian, \(\eta_p\ket{G}=0\) although \(\bra{G}\bar{\eta}_p\ne 0\). As discussed in Refs. \cite{lee_prx_2014, turkeshi_prb_2023, turkeshi_prb_2023b}, we observe \(\bra{G}\eta_p^{ \dagger}=0\)
. More specifically, we define the ground state as the right eigenvector, denoted by $\ket{R}$, corresponding to the eigenvalue with the smallest imaginary part in the broken regime while in the unbroken regime, it is well-defined through the lowest real eigenvalue. The state of the system is then described by the normalized steady state under post-selection, given by $\rho=\frac{\ket{R}\bra{R}}{\text{Tr}(\ket{R}\bra{R})}$. This framework is commonly used to analyze the properties of non-Hermitian Hamiltonians.

In this study, we compute the entanglement between nearest-neighbor  sites, $i$ and $i+1$, in a non-Hermitian system, which can be obtained by tracing out $(N-2)$ parties of the $N-$party ground state. Note here that any two-qubit density matrix between site $i$ and $j$, denoted as $\rho_{ij}$, can be written as
\begin{equation}
    \rho_{ij}=\frac{1}{4} \left (\mathbb{I}_4 + \vec{r} \cdot \vec{\sigma}  \otimes \mathbb{I} + \mathbb{I} \otimes \vec{s} \cdot \vec{\sigma} + \sum_{k,l = x,y,z} C^{kl}_{ij} \sigma^{k}_i \otimes \sigma^{l}_j\right),
\end{equation}
where the coefficients of Pauli matrices, $r_i^k = \text{Tr}(\sigma^k_i \rho_i)$ and $s_j^k = \text{Tr}(\sigma^k_j \rho_j) $ determine the magnetization in the \(x,y,\, \text{and}\, z\)-directions, with $\rho_i$ being the reduced single-site density matrix of $\rho_{ij}$ and \(C^{kl}_{ij} = \text{Tr}(\sigma_i^k \otimes \sigma_j^l \rho_{ij})\) denotes the classical correlators. Due to Wick's theorem, an unbalanced number of fermionic creation and annihilation operators is in $m^x$, $m^y$, $C^{yz}$, $C^{zy}$, $C^{zx}$ and $C^{xz}$ and they all vanish $\forall$ $i$ and $j$. Hence  \(\rho_{ij}\) can be expressed as
\begin{equation}
    \rho_{ij}=\frac{1}{4}\! \left (\mathbb{I}_4 + m^z({\sigma^z_j}\!+\!{\sigma^z_i}) + C^{zz}_{ij} \sigma^{z}_i\!\! \otimes\! \sigma^{z}_j+ \sum_{k,l=x,y}\!\!\! C^{kl}_{ij} \sigma^{k}_i \!\!\otimes \!\sigma^{l}_j\right).
    \label{eq:desnity_matrix}
\end{equation}
Now, by expressing \(a_p\) and \(a_p^\dagger\) in terms of \(\eta_p\) and \(\eta_p^\dagger\), we have 
\begin{equation}
    a_p=-\frac{v_p^{1*}\eta_{p}+u_p\eta_{-p}^\dagger}{\abs{u_p}^2+\abs{v_p^1}^2} \quad \text{ and }\quad a_{-p}=\frac{v_p^{1*}\eta_{-p}-u_p\eta_{p}^\dagger}{\abs{u_p}^2+\abs{v_p^1}^2}.
\end{equation}
Note that \(\eta_p \text{ and } \eta_p^\dagger\) do not obey fermionic anti-commutation rule, and \(\expval{\eta_p\eta_p^\dagger}=\abs{v_p^1}^2+\abs{u_p}^2\), where the expectation value is calculated with respect to the ground state. In the thermodynamic limit, all the non-vanishing classical correlators and magnetization involved in the two-site density matrix of the ground state can be obtained in terms of $u_p$ and $v_p^i (i = 1,2)$ and are given by 
\begin{eqnarray}
    m^z &=& \frac{1}{\pi}\int_{0}^{\pi}d\phi_p \Lambda_p,\nonumber\\
    C^{xy} &=& C^{yx}= \frac{1}{\pi}\int_{0}^{\pi} d\phi_p \Omega^+_p\sin \phi_p, \nonumber\\
    C^{xx} &=& \frac{i}{\pi}\int_{0}^{\pi} d\phi_p \Omega^-_p\sin \phi_p + \frac{1}{\pi}\int_{0}^{\pi}d\phi_p\Lambda_p\cos \phi_p, \nonumber\\
    C^{yy} &=&-\frac{i}{\pi}\int_{0}^{\pi} d\phi_p \Omega^-_p\sin \phi_p + \frac{1}{\pi}\int_{0}^{\pi}d\phi_p \Lambda_p\cos \phi_p, \nonumber\\
    \text{ and } \quad C^{zz} &=& (m^z)^2-C^{xx}C^{yy}+C^{xy}C^{yx}.
    \label{eq:corr}
\end{eqnarray}
Here $\Lambda_p = \frac{\abs{v_p^1}^2-\abs{u_p}^2}{\abs{v_p^1}^2+\abs{u_p}^2}$, $\Omega^-_p =\frac{u_p^*v_p^1-v_p^{1*}u_p}{\abs{v_p^1}^2+\abs{u_p}^2}$ and $\Omega^+_p =\frac{v_p^{1*}u_p+u_p^*v_p^1}{\abs{v_p^1}^2+\abs{u_p}^2}$. Therefore, we can now compute $\rho_{i\:i+1}$ and its entanglement content in terms of logarithmic negativity (see Appendix. \ref{app:log_neg}) which can help us to probe Region-I and Region-II of the $H^{\text{iKSEA}}_p$ model. 

\begin{figure}
     \hspace*{-0.6cm}
    \includegraphics[width=\linewidth]{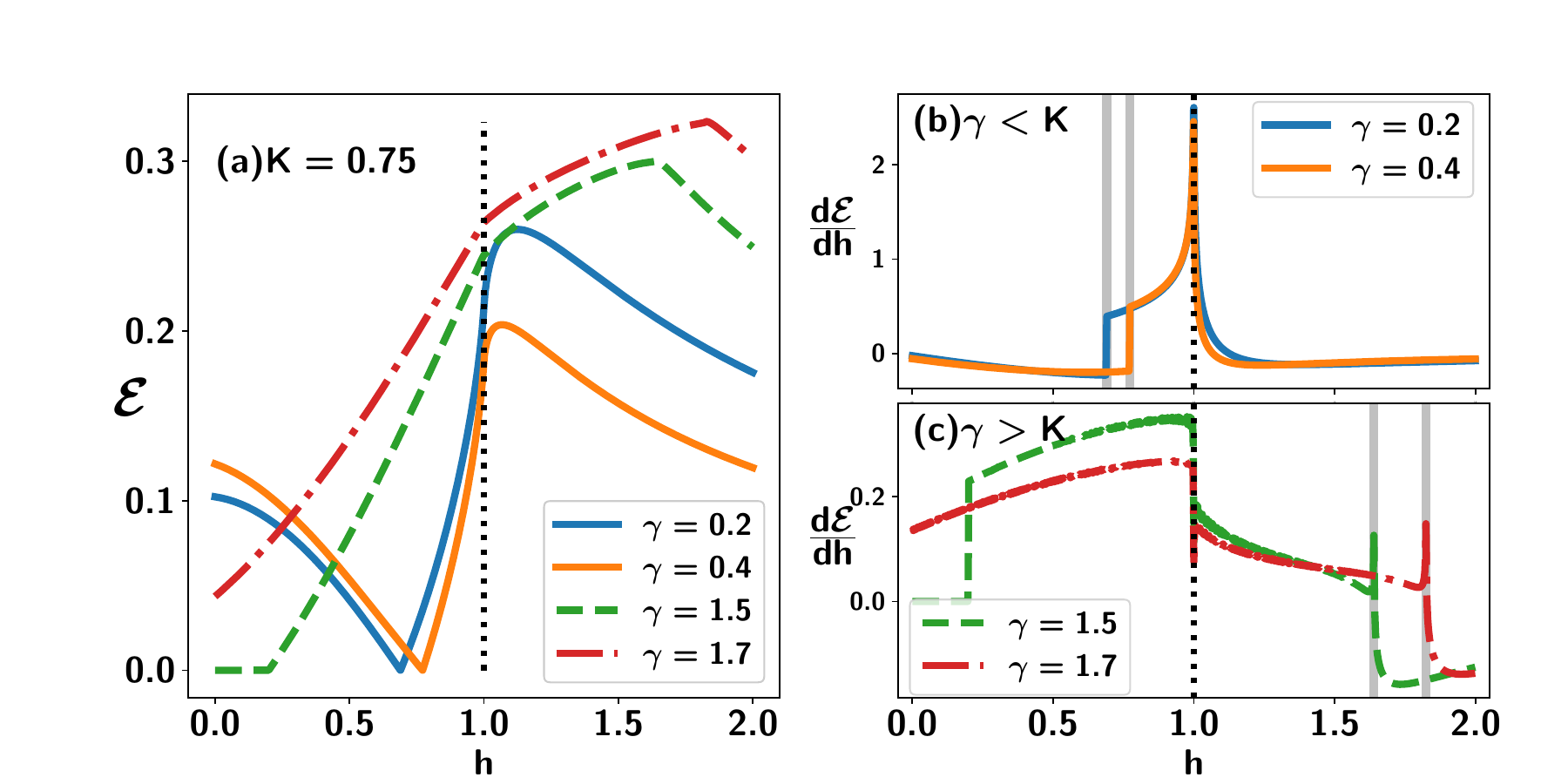}
    \caption{ {\bf Variation of nearest-neighbor entanglement, \(\mathcal{E}\) (ordinate) as a function of external magnetic field, \(h\) (abscissa).} (a)  Bipartite entanglement between the nearest neighbor spins for $\gamma<K$ (solid lines) and $\gamma>K$ (dashed  and dashed-dotted lines). Other system parameters are $K = 0.75$ and \(N=5000\).  (b) Derivative of nearest-neighbor entanglement  with \(h\) for \(\gamma<K\). \(\frac{d\mathcal{E}}{dh}\) is non-analytic at the factorization points (gray vertical lines) and at the critical point $h_c=1$ (dotted vertical line). (c) \(\frac{d\mathcal{E}}{dh}\) vs \(h\) for \(\gamma>K\) which is non-analytic at the exceptional points (gray vertical lines) and at the critical point, $h_c=1$ (dotted vertical line). All the axes are dimensionless.}
    \label{fig:dervative_ent}
\end{figure}

\textbf{Region-I ($\gamma > K$).} In this scenario, both the broken and the unbroken phases exist in the $(h, \gamma)$-plane as discussed before. The ground state is unique in the unbroken phase and is taken as the tensor product of all eigenvectors with the lowest eigenvalues of each momentum block. On the other hand, in the $\mathcal{RT}$-symmetry broken regime, the definition of the ground state is ambiguous as the imaginary eigenvalues have no definite ordering. However, the choice of the ground state is made where the eigenvector corresponds to the eigenvalue with the lowest imaginary term. With this convention at hand, we compute the nearest-neighbor entanglement, denoted by \(\mathcal{E}\) quantified via logarithmic negativity \cite{vidal_pra_2002}, of the ground state by varying both $h$ and $\gamma$ as depicted in Fig. \ref{fig:ent2}. The $(\gamma,h)$-plane in this region contains two transition points : $(1)$ exceptional point, where eigenvectors coalesce and cannot exist in Hermitian models, and (2) quantum critical point in which gap-closing occurs. Interestingly, non-vanishing bipartite entanglement changes its curvature (see Fig. \ref{fig:dervative_ent}(a)) at both the points. Specifically, for a fixed non-Hermiticity parameter, which dominates over the Hermitian KSEA interaction, the first derivative of bipartite entanglement with respect to the magnetic field shows nonanalytic behavior at both the critical and exceptional points (as shown in Fig. \ref{fig:dervative_ent}(c)) present in the non-Hermitian models.

\textbf{Region-II ($\gamma\leq K$).} The ground state in this domain possesses two interesting characteristics with the variation of $\gamma$ and $h$ -- (1) critical point at $h_c = 1$ and (2) factorization surface. Notice that the system is quite similar to the transverse $XY$ model in the Hermitian case, although there are some stark differences. Specifically, for a fixed non-Hermiticity parameter, bipartite entanglement first decreases with the change of the magnetic field, vanishes at the factorization surface, $h_f = \sqrt{1+\gamma^2-K^2}$ and then starts increasing with the increase of $h$ until it reaches its maximal value. Importantly, $\pdv{\mathcal{E}}{h}$ again shows a kink (non-analyticity) at the gap-closing point $h_c = 1$. 

{\it Factorization surface.} Here we report the factorization surface for the non-Hermitian model, which is well-studied in the Hermitian system \cite{giampaolo_prl_2008,giampaolo_prb_2009}. In particular, we find $\mathcal{E}=0$ at $h_f = \sqrt{1+\gamma^2-K^2}$, which is different from the factorization surface, known in the Hermitian model, given by $h_f^{\text{Herm}} = \sqrt{1-\gamma^2-K^2}$. Interestingly, when $K=0$, the non-Hermitian $iXY$ model in the presence of uniform as well as alternating magnetic fields does not contain factorization surface, thereby illustrating the significance of having KSEA interaction in the non-Hermitian system. Moreover, at $\gamma = K$, $\mathcal{E}=0$ $\forall~h \ge 1$ (see Fig. \ref{fig:dervative_ent}(a) and (b)).

The behavior of the nearest-neighbor entanglement in this region can be explained from another perspective. Let us consider a gauge transformation of the fermionic operators such that
\begin{equation}
    c_j= e^{\mu/2}e^{i\theta/2}c_j, \quad \text{and} \quad \bar{c}_j= e^{-\mu/2}e^{-i\theta/2}c_j^\dagger,\
    \label{eq:map_xy_ksea}
\end{equation}
where \(e^{\mu}=\sqrt{\frac{K-\gamma}{K+\gamma}}\) and \(\theta=-\pi/2\) with \(\{c_j,\bar{c}_k\}=\delta_{jk}\). According to this transformation, we can re-express Eq. (\ref{eq:iksea_jw}) as

\begin{eqnarray}
    \nonumber H^{XY}_{JW}&=&\frac{1}{2}\sum_j (\bar{c}_jc_{j+1}+\bar{c}_{j+1} c_{j})+ \gamma'(\bar{c}_j \bar{c}_{j+1}\\&&+c_{j+1} c_j)+h(2\bar{c}_j c_{j}-1),
    \label{eq:xy_jw}
\end{eqnarray}
where \(\gamma'=\sqrt{K^2-\gamma^2}\). This is the free-fermionic version of the Hermitian \(XY\) spin model with transverse magnetic field. The factorization surface of the \(XY\)-spin model, \(h_f^{XY}=\sqrt{1-\gamma'^2}\) \cite{giampaolo_prl_2008, giampaolo_prb_2009}. Interestingly, it coincides with the factorization surface,  \(h_f=\sqrt{1+\gamma^2-K^2}\) of the non-Hermitian iKSEA in the unbroken region. By the mapping in Eq. (\ref{eq:map_xy_ksea}), we can obtain a justification for the similar behavior of nearest-neighbor entanglement in the iKSEA model in this unbroken region and the transverse $XY$ model. The entire analysis establishes that even for the non-Hermitian system, bipartite entanglement can be a good indicator to identify all kinds of transitions, ranging from gap-closing of the energy spectrum, coalescence of eigenvectors, to vanishing entanglement surface. More ambitiously, in the succeeding section, we probe how entanglement is capable of detecting these transitions even under quenching. Note that in literature, the block entanglement entropy of the entire system is explored to calibrate properties of the eigenstates of a non-Hermitian Hamiltonian which is different from our cases~\cite{kawabata_prx_2023,gal_scipost_2023, chen_prb_2022}. 

\begin{figure}
     \centering
    \includegraphics[width=\linewidth]{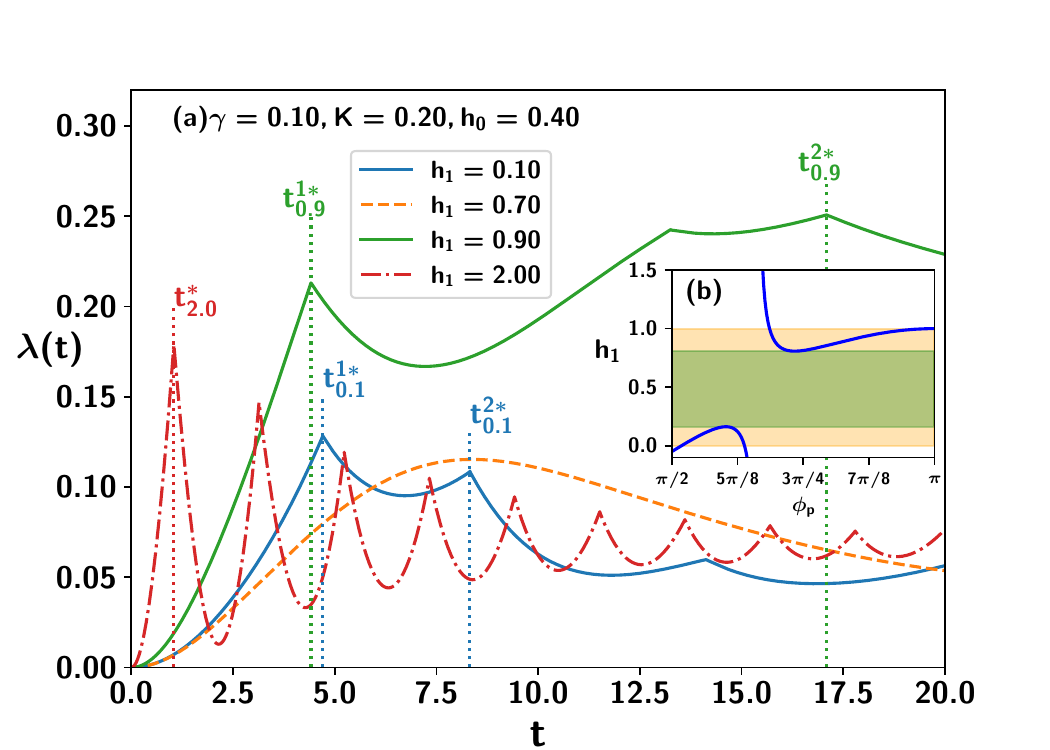}
    \caption{Rate function, \(\lambda(t)\) (vertical axis) as a function of time, \(t\) (horizontal axis). (a) Critical times are mentioned in the graph as \(t^{*}_{h_1}\). The system size is \(N=5000\) and other parameters are $\gamma=0.1, K=0.2$, \(h_0=0.4\). Different lines represents \(\lambda(t)\) for quenching in different $h_1$ values, with $h_1=0.1, 0.9$ and $2.0$ showing DQPT. (b) DQPT condition on quenched parameter, $h_1$ (solid blue curve) according to Eq.(\ref{eq:h1_cond}). No DQPT occurs when the value of $h_1$ lies in the green (dark shaded) region $h_1\in (h_a, h_b)$, while when it crosses the yellow (light shaded) region $h_1\in[0, h_a]\cup[h_b,1]$, it shows DQPT without crossing the critical lines. Here $h_a\sim0.1628, h_b\sim0.8072$ and all the axes are dimensionless.}
    \label{fig:dqpt}
\end{figure}

\section{Dynamical behavior: Loschmidt echo and Entanglement}
\label{sec:dynamics}

Non-Hermitian systems naturally occur under dissipative dynamics of a larger Hermitian systems undergoing unitary evolution. Thus it is imperative to study the dynamical behavior of the physical quantities under non-Hermitian evolution. For carrying out such investigations, the set up considered here is as follows: 

\noindent (1) \emph{Initial state preparation.} - The ground state of the Hamiltonian \(H^{\text{iKSEA}}_p(h_0)=H_p^0\) with magnetic field, \(h_0\), is chosen to be the initial state which is taken to be the right eigenvector corresponding to the ground state energy,  written as \(\ket{\psi'(0)_p}=\left( -v_p^1, u_p\right)^T\). In order to detect probability or construct density matrix, the normalization of the eigenstate of \(H_p^0\) is performed using the standard Dirac inner product \cite{lin_jopa_2016, yang_pra_2020}, i.e, we should consider
\begin{equation}
    \ket{\psi(0)_p}=\frac{\ket{\psi'(0)_p}}{\sqrt{\abs{v_p^1}^2+\abs{u_p}^2}}.
\end{equation}
(2) \emph{Dynamics.} - The initial magnetic field, \(h_0\), is suddenly quenched to \(h_1\) at \(t>0\), which leads to the evolution of the system governed by the Hamiltonian, \(H^{\text{iKSEA}}_p(h_1)=H_p^1\). The time evolved state is written in the momentum block as
\begin{equation}
    \ket{\psi(t)^\prime_p}\!=\!e^{-iH_p^1t}\ket{\psi(0)_p};\:\: \ket{\psi(t)_p}\!=\!\frac{\ket{\psi(t)^\prime_p}}{||\ket{\psi(t)^\prime_p}||}=
    \begin{bmatrix}
    -v^1_p(t) \\
    u_p(t)
    \end{bmatrix},
    \label{eq:evol_state}
\end{equation}
which leads to the evolved state of the system as \(\ket{\Psi(t)}=\otimes_p\ket{\Psi(t)_p}\). The idea is to choose \(h_0\) and \(h_1\) across transition lines and study the effects on dynamical quantities like Loschmidt echo, the corresponding rate function, and entanglement.

\subsection{Mimicking criticalities via rate function and its breakdown}
\label{sec:losh_echo}
To analyze whether the dynamical properties of the system can reflect the previously established criticalities, the initial state preparation and the quenching are performed in Region - II with $\gamma \le K$. Before reporting the results, let us define the dynamical quantity, the Loschmidt echo (LE), and the corresponding rate function as 
\begin{equation}
    \mathcal{L}(t) = \frac{\norm{\langle\Psi(0)|\Psi(t)\rangle}^2}{\norm{\langle\Psi(t)|\Psi(t)\rangle}}, \quad  \quad \lambda(t) = -\frac{1}{N}\log{\mathcal{L}(t)}.
\end{equation}
Both quantities calculated in the dynamics correspond to dynamical free energy, and their non-analyticities with time have been shown to signal equilibrium quantum phase transition efficiently for Hermitian systems 
\cite{heyl_prl_2013, Heyl_review_2018} (for exceptions, see Refs.~\cite{Vajna14, Sirker14, gurarie_pra_2019, stav_prb_2020}). 


The Loschmidt echo (LE) for the iKSEA model takes the form 
\begin{equation}
    \mathcal{L}(t)=\prod_p \norm{\cos\epsilon_p^1 t - i\bra{\psi(0)}H^1_p\ket{\psi(0)}\frac{\sin\epsilon_p^1 t}{\epsilon_p^1}}^2,
\end{equation}
where $\epsilon_p^1 $ is the energy of the quenched Hamiltonian, $H_p^1$. First note that LE vanishes, when $\epsilon_p^1$ is only real, i.e., when the quenching is done by the Hamiltonian in the unbroken phase. This keeps both the $\cos\epsilon_p^1 t$ and the $\frac{\sin\epsilon_p^1 t}{\epsilon_p^1}$ terms always real. 

In the thermodynamic limit, there exists a momentum, $\phi_{p_c}\in(0,\pi)$ such that $\bra{\psi(0)}H^1_{p_c}\ket{\psi(0)}=0$, which provides a necessary and sufficient condition for LE to become zero and implies that DQPT occurs. Preparing the ground state of the Hamiltonian with the local magnetic field, $h_0$ and quenching to $h_1$, the relation in terms of $\gamma$ and $K$ ($\gamma<K$) is given by
\begin{equation}
    h_1 = h_0 + \epsilon_{p_c}^0\frac{(h_0+\cos\phi_{p_c}-\epsilon_{p_c}^0)^2+(\gamma+K)^2\sin^2\phi_{p_c}}{(h_0+\cos\phi_{p_c}-\epsilon_{p_c}^0)^2-(\gamma+K)^2\sin^2\phi_{p_c}},
    \label{eq:h1_cond}
\end{equation}
where \(\epsilon_{p_c}^0\) is the initial energy at the critical momenta. This leads to the critical time for DQPT as $t_n^*=\frac{\pi}{\epsilon_{p_c}^1}(n+\frac{1}{2})$, $ (n \in \mathrm{Z})$, where $\epsilon_{p_c}^1 = \sqrt{(h_1+\cos\phi_{p_c})^2-(\gamma^2-K^2)\sin^2\phi_{p_c}} $.


\begin{figure}
     \centering
    \includegraphics[scale=0.35]{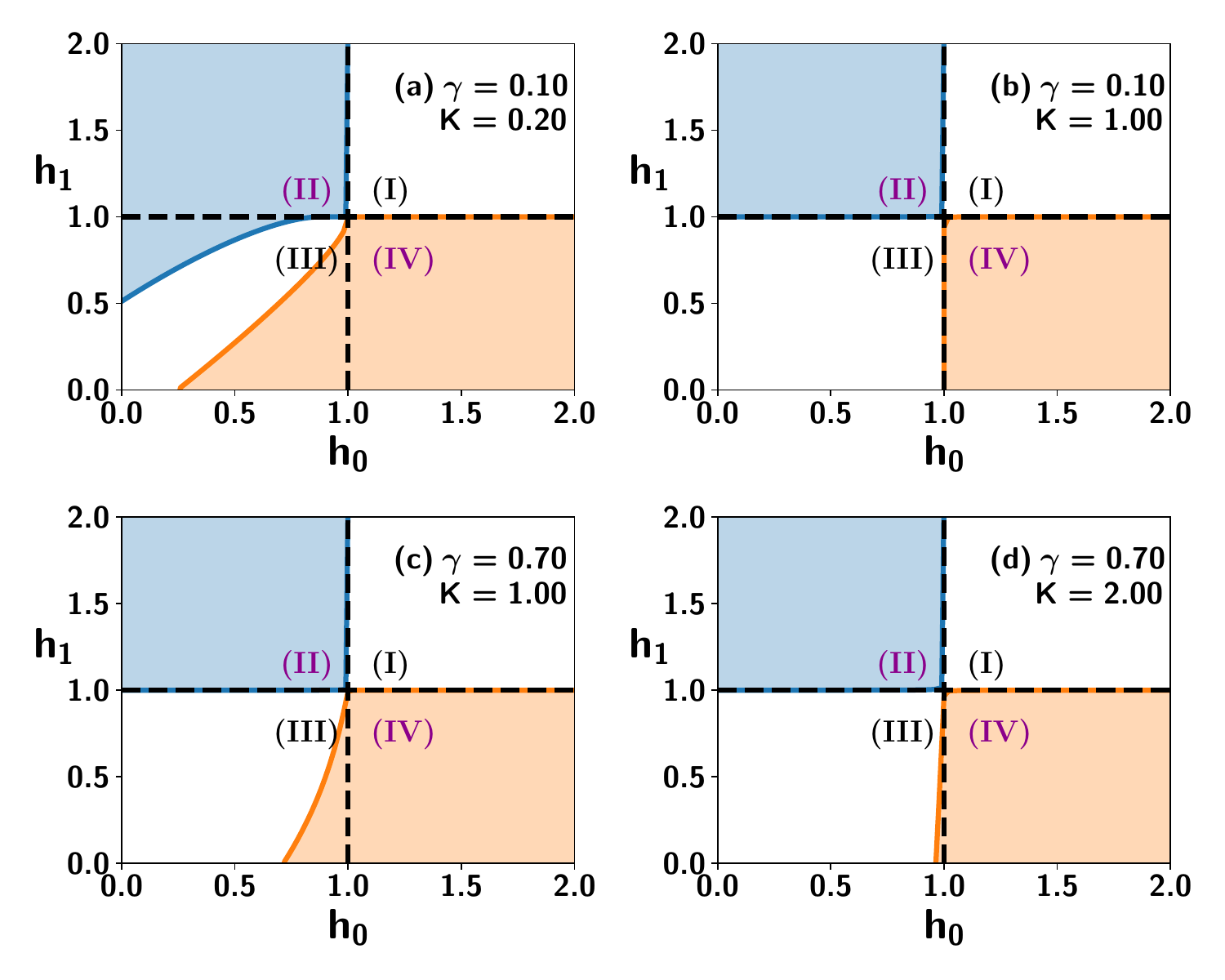}
    \caption{{\bf Non-analyticities of the rate function in the (\(h_0, h_1\))-plane} for different pairs of \(\gamma\) and \(K\) values. The entire region is divided into four quadrants -- QI and QIII corresponds to initial and quenching magnetic fields without crossing the critical line $h_c=1$, while the quadrants, QII and QIV requires the crossing of the critical line. Rate functions correctly detect QII and QIV regions, thereby efficiently identifying DQPT.  However, it is not able to signal the quantum phase transition present in equilibrium since non-analytic behavior is observed even when the initial and final parameters are chosen from the same phase,  i.e., when both $0<h_0<1$ and $0<h_1<1$ in QIII. Notice, however, that when both the initial and quenched magnetic fields are above unity, the rate function does not show any non-analyticity. 
    Other parameters of the system are presented in the legend. All axes are dimensionless.}
    \label{fig:dqpt_detect}
\end{figure}
This shows that when the initial and quenching Hamiltonian is taken with $h_0$ and $h_1$ respectively having $\gamma \le K$, DQPT occurs if and only if there exists a critical momentum $\phi_{p_c}\in(0,\pi)$, so that $h_1$ satisfies Eq. (\ref{eq:h1_cond}). Using this condition, we prove that LE obtained via quenching dynamics cannot always mimic the transition which is present as shown in Fig. \ref{fig:dqpt}(a). 

To examine more precisely, we analyze for several initial states and final quenches in the $(h_0, h_1)$-plane for a fixed $\gamma$ and $K$. Specifically, we divide the entire $(h_0,h_1)$-plane in four quadrants, namely QI ($1<h_0<2, 1<h_1<2$), QII ($0<h_0<1, h_1>1$), QIII ($0<h_0<1, 0<h_1<1$), and QIV ($h_0>1, 0<h_1<1$) for identifying $h_c = 1$ in dynamics. The quadrants I and III correspond to quench in which no critical line is crossed for initial and quenched Hamiltonian, while the other two quadrants represent the scenario with the initial and the final states being taken across the critical lines. For an ideal DQPT signature, the rate function should not be non-analytic in the QI and QIII regions. However, in Fig. \ref{fig:dqpt_detect}, this is not the case. Although the rate function can faithfully distinguish the situation when the quenching is performed across the critical line as depicted for QII and QIV, there are pairs of $(K,\gamma)$ values where non-analyticities in the rate function are observed in QIII. Another interesting finding is that the rate function never becomes non-analytic when both the magnetic fields, $h_0$ and $h_1$ are strong i.e., $(h_0,h_1) > h_c\equiv 1$. Moreover, we find that there are two critical momenta which correspond to two different critical times when $0<h_0<1$ and $0<h_1<1$ while a unique time is obtained when $h_0 < 1$ and $h_1 > 1$ (QII) or $h_0 > 1$ and $h_1 < 1$ (QIV) (as shown in Fig. \ref{fig:dqpt}(b)). This is the reason behind the non-analytic behavior of the rate function seen in different quadrants including the false signature. 

\begin{figure}
     \centering
    \includegraphics[width=\linewidth]{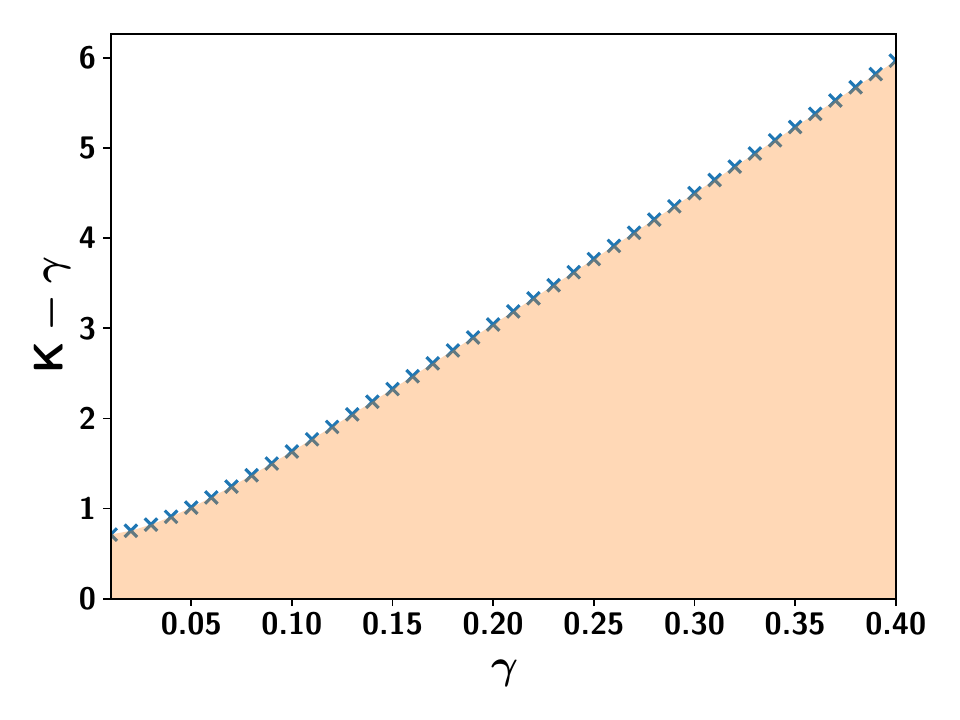}
    \caption{Conditions on $K$ and $\gamma$ for dynamical phase transition mimicking the critical lines. Crosses separate the region in two sectors -- shaded portion corresponds to the region in which non-analyticity in the rate function incorrectly  identify the quenching and the initial magnetic field regimes (same as QIII in Fig. \ref{fig:dqpt_detect}) while the rate function detects DQPT in the white portion only when the critical line is crossed. 
    Both the axes are dimensionless. 
    }
    \label{fig:dqpt_params}
\end{figure}

After visualizing the picture with respect to the variation of $h_0$ and $h_1$, let us focus on the role of $\gamma$ and $K$ in finding non-anayticities in the rate function with respect to the quenching in the magnetic field. Note that the values of $\gamma$ represent the amount of non-Hermiticity in the model, while with high values of $K$, the system approaches the Hermitian one. Our careful analysis reveals that the wrong signal from the rate function is obtained when $\gamma$ dominates over $K$. The rate function can correctly signal the crossing of critical lines during the quench when the difference between $K$ and $\gamma$ increases (see Fig. (\ref{fig:dqpt_params}) for depiction).

\begin{figure}
     \centering
    \includegraphics[width=\linewidth]{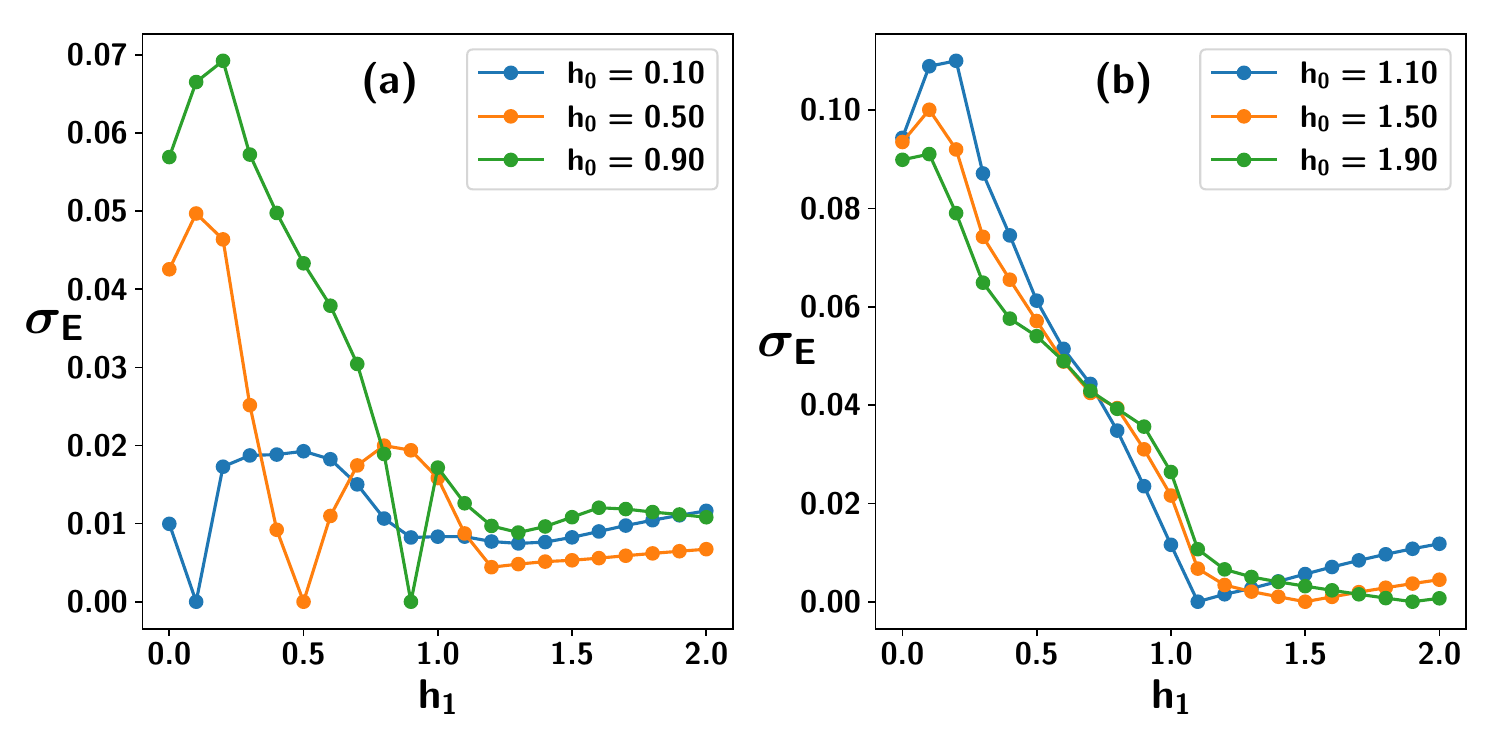}
    \caption{Fluctuations in the dynamics of nearest-neighbor entanglement (ordinate), quantified via \(\sigma_E\), defined in Eq. (\ref{eq:entsecond}) against the strength of the magnetic field in the quenching Hamiltonian, \(h_1\) (abscissa). 
    Here $\gamma=0.5, K=1$, with $N=2000$. 
    (a). The initial magnetic fields are chosen below the critical line. Note that the left portion where \(h_1<1\) corresponds to the QIII while the parameters belong to QII when \(h_1>1\).  (b). \(h_0\) is below the critical point, \(h_c=1\) and it covers the QI and QIV. Both the axes are dimensionless.  }
    \label{fig:ent_dyn1}
\end{figure}

\begin{figure}
     \centering
    \includegraphics[width=\linewidth]{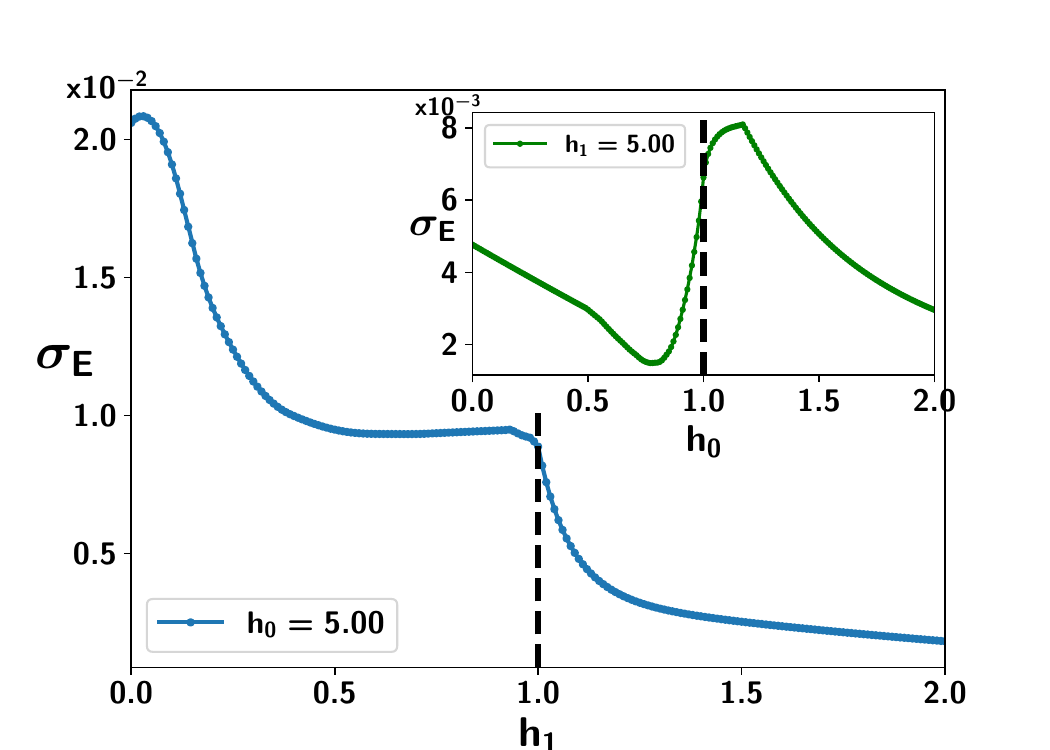}
    \caption{\(\sigma_E\) (ordinate) vs \(h_1\) (abscissa) for a fixed value of \(h_0=5\). All other specifications are same as in Fig. \ref{fig:ent_dyn1}. (Inset) It is similar to the main figure except we interchange  \(h_0\) and \(h_1\), i.e., the quenching magnetic field is fixed to \(h_1=5\) and the behavior of \(\sigma_E\) is plotted with respect to \(h_0\). Clearly, \(\sigma_E\) determines the critical point, \(h_c=1\). 
    Both the axes are dimensionless.  }
    \label{fig:ent_dyn2}
\end{figure}

\subsection{Determination of criticalities via fluctuation of dynamical entanglement}
\label{sec:dyn_ent}
In Hermitian systems, nearest-neighbor entanglement is shown to be capable of detecting the equilibrium phase transition occurring at zero temperature \cite{ fazio_nature_2002, osborne_pra_2002} and the dynamical quantum phase transition \cite{fazio_review_2008, Aditi2005, sanpera_chiara_2018, nielsen_pra_2011}. Let us examine how bipartite entanglement is effective in the case of the non-Hermitian model. Without evolution, we have already shown in Sec. \ref{sec:static_entang} that the critical points including exceptional points can be identified via the patterns of bipartite entanglement. Even though the evolution is non-Hermitian, it takes the Gaussian state to another Gaussian state, hence, we can easily characterize the two-party state with time-dependent coefficients $u_p(t)$ and $v_p^1(t)$ in Eq.~(\ref{eq:evol_state}), which makes the $\Lambda_p(t)$ and $\Omega^{\pm}_p(t)$ time-dependent. Therefore, the two-point correlations at each time can easily be obtained from Eq.~(\ref{eq:corr}) as in the equilibrium case, which, in turn, helps to write the time-dependent two-party density matrix using Eq.~(\ref{eq:desnity_matrix}). We are interested to study the behavior of nearest-neighbor entanglement $\mathcal{E}(t)$ with variation of time. We observe that the entanglement fluctuation pattern can discriminate the quadrants mentioned in Fig. \ref{fig:dqpt_detect} for a fixed non-Hermiticity parameter $\gamma$ and the KSEA strength $K$. To quantify such an effect, we introduce a quantity 
\begin{equation}    \sigma_{\mathcal{E}}=\lim_{t\to\infty}\sqrt{\expval{\mathcal{E}(t)^2}_t-\expval{\mathcal{E}(t)}_t},
\label{eq:entsecond}
\end{equation} 
as a figure of merit to detect quantum critical points using dynamical entanglement. 
Let us summarize the result according to four quadrants in Fig. \ref{fig:dqpt_detect}. In particular, in the first quadrant QI, $\sigma_{\mathcal{E}}$ is very low and is not sensitive to the change of initial magnetic field, $h_0$ while when the system parameters are tuned in QIV, $\sigma_{\mathcal{E}}$ is comparatively high although it does not vary with the initial magnetic field $h_0$ (as shown in Fig. \ref{fig:ent_dyn1}(b)). On the other hand, $\sigma_{\mathcal{E}}$ is extremely sensitive with the choice of the initial magnetic field $h_0$ in the QIII
region.  In QII, the fluctuations of entanglement are small, irrespective of the values of $h_0$ (as shown in Fig. \ref{fig:ent_dyn1}(a)).  
The signature of criticality from $\sigma_{\mathcal{E}}$ becomes prominent when $h_0 >> h_c$ or $h_1>> h_c$, i.e., in the domain of QI and QIV (see Fig. \ref{fig:ent_dyn2}). 
In particular, when $h_1<h_c$, $\sigma_{\mathcal{E}}$ is high and then saturates near $h_c = 1$ while it suddenly changes its curvature when $h_1 > h_c$ with \(h_0 >>h_c\). Similar behavior also emerges when the final quenching magnetic field is also very strong and the initial field is varying, as depicted in the inset of Fig. \ref{fig:ent_dyn2}.


\section{Conclusion}
\label{sec:conclu}

Non-Hermitian many-body systems which can be obtained from Hermitian systems connected with an environment, can reveal some exquisite features, typically absent in their Hermitian counterparts. In particular, along with the critical points where the gap between eigenenergies vanishes, the non-Hermiticity leads to an exceptional point where eigenvectors coalesce and it divides the system into two phases, broken and unbroken regimes. 

To explore the non-Hermitian system, we investigated a system containing \(\mathcal{RT}\)-symmetric \(iXY\) having imaginary non-Hermiticity parameter and Hermitian Kaplan-Shekhtman-Entin-Aharony (KSEA) interactions along with the transverse magnetic field. We realize that when a $XX$ spin chain with KSEA interaction is in contact with local and non-local baths, the effective Hamiltonian represents the non-Hermitian model studied in this work.
We found that there is a trade-off relation between the strength of non-Hermiticity and Hermitian KSEA interaction strength. In particular, by exploiting the Jordan-Wigner transformation, we identified that the system possesses three interesting transitions -- exceptional points dividing the broken-unbroken region, factorization point having vanishing entanglement in the ground state, and the critical point where the energy gap vanishes in the transverse magnetic field line. We report that the nearest-neighbor entanglement can successfully determine all these points. To our knowledge, the factorization surface in the non-Hermitian model has never been determined in the literature.

Starting from the ground state of the model, we observed that both the rate function and the bipartite entanglement can capture the situation when the quenching Hamiltonian is in a domain different from the initial Hamiltonian. However, there are regimes where the rate function can signal the criticality incorrectly. In the case of bipartite entanglement in the evolved state, we found that the fluctuations quantified by the second moment of nearest-neighbor entanglement after averaging over a long time can accurately predict the criticalities present in the model. Our investigations establish that entanglement is competent to determine both the Hermitian and non-Hermitian properties of a many-body system. 


\section*{Acknowledgements}

We acknowledge the support from Interdisciplinary Cyber Physical Systems (ICPS) program of the Department of Science and Technology (DST), India, Grant No.: DST/ICPS/QuST/Theme- 1/2019/23. We  acknowledge the use of \href{https://github.com/titaschanda/QIClib}{QIClib} -- a modern C++ library for general purpose quantum information processing and quantum computing~\cite{qiclib} and cluster computing facility at Harish-Chandra Research Institute. L.G.C.L. is funded by the European Union. Views and opinions expressed are, however, those of the author(s)
only and do not necessarily reflect those of the European Union or the European Commission. Neither the
European Union nor the granting authority can be held responsible for them. This project has been funded
by the Caritro Foundation. This work was supported by the Provincia Autonoma di Trento, and Q@TN, the
joint lab between the University of Trento, FBK—Fondazione Bruno Kessler, INFN—National Institute for
Nuclear Physics, and CNR—National Research Council, Italy.
\appendix

\section{Logarithmic negativity}
\label{app:log_neg}
 We quantify entanglement in two-qubit systems by logarithmic-negativity (LN). For a given bipartite state \(\rho_{ij}\) between two site, \(i\) and \(j\), LN \cite{vidal_pra_2002} is defined as $$\mathcal{E}(\rho_{ij}) = \log_2(2\mathcal{N}(\rho_{ij})+1),$$ where $\mathcal{N}$ is the absolute sum of negative eigenvalues of the partially transposed state ($\rho_{ij}^{T_i}$ or $\rho_{ij}^{T_j}$) with partial transposition being taken with respect to either \(i\) or \(j\) \cite{peres_prl_1996,horodecki_pla_1996}. Since \(\rho_{ij}\) is a two-qubit state, $\mathcal{E}(\rho_{ij})=0$ is a necessary and sufficient condition for the state to be separable in this work \cite{horodecki_pla_1996}.
 
\section{Factorization surface of KSEA model}
\label{sec: factorization_KSEA}
Let us derive the factorization surface for the Hermitian KSEA model which is described as
\begin{eqnarray}
    \nonumber H^{\text{KSEA}}&=&\sum_{j}\frac{(1+\gamma)}{4}\sigma_j^x\sigma_{j+1}^x+\frac{(1-\gamma)}{4}\sigma_j^y\sigma_{j+1}^y\\&&+\frac{K}{4}(\sigma_j^x\sigma_{j+1}^y+\sigma_j^y\sigma_{j+1}^x)+\frac{h}{2}\sigma_j^z,
\end{eqnarray}
where \(\sigma_{N+1}=\sigma_1\). Due to periodic boundary condition(PBC), the separable state at the factorization point can be written as
\begin{equation}
    \ket{\Phi}=\prod_{j=0}^{N/2-1} \ket{\phi_{2j}^e}\otimes\ket{\phi_{2j+1}^o},
\end{equation}
where \(\ket{\phi^{(o)e}}\) are the state at odd (even) site. The Hamiltonian can be written by dividing into odd and even sector as \(H^{\text{KSEA}}=\sum_{j}^{N/2-1}(H_{2j,2j+1}^{eo}+H_{2j+1,2j+2}^{oe})\) where \(H^{eo}\) is a two-site Hamiltonian given by the form,
\begin{eqnarray}
    \nonumber H^{eo}&=&\frac{(1+\gamma)}{4}\sigma_e^x\sigma_o^x+\frac{(1-\gamma)}{4}\sigma_e^y\sigma_{o}^y+\frac{h}{4}\sigma_e^z+\frac{h}{4}\sigma_o^z\\&&+\frac{K}{4}(\sigma_e^x\sigma_{o}^y+\sigma_e^y\sigma_{o}^x).
\end{eqnarray}
We have to show that \(\ket{\Phi}\) is the eigenstate of the \(H^{\text{KSEA}}\) corresponding to the minimum energy and the parameter space has to be identified which leads to the ground state \(H^{\text{KSEA}}\). Let us compute the minimum energy obtained from \(H^{\text{KSEA}}\) via \(H^{\text{KSEA}}\) where the minimization is performed over the set of product states, i.e,
\begin{eqnarray}
  \nonumber E_{min}&=&\underset{\ket{\phi^e}\ket{\phi^o}}{\min}\bra{\Phi}H^{\text{KSEA}}\ket{\Phi}\\&=&N\underset{\ket{\phi^e}\ket{\phi^o}}{\min}\bra{\phi^e}\bra{\phi^o}H^{eo}\ket{\phi^e}\ket{\phi^o},
\end{eqnarray}
where we have used the fact that \(H^{eo}\) and \(H^{oe}\) are energetically equivalent. Hence, the minimization depends upon the separable energy per site which is \(\delta=\underset{\ket{\phi^e}\ket{\phi^o}}{\min}\bra{\phi^e}\bra{\phi^o}H^{eo}\ket{\phi^e}\ket{\phi^o}\) and the states, \(\ket{\phi^{(o)e}}\) is, in general can be written as
\begin{equation}
    \ket{\phi^{(o)e}}=\cos\frac{\alpha^{e(o)}}{2}\ket{0}+\exp (i\beta^{e(o)})\sin\frac{\alpha^{e(o)}}{2}\ket{1},
\end{equation}
where \(0\le \alpha^{e(o)}\le \pi\) and \(0\le \beta^{e(o)}\le 2\pi\). The minimum energy, \(\delta\) is re-expressed as
\begin{widetext}
\begin{equation}
    \delta=\underset{\alpha^{e(o)},\beta^{e(o)}}{\min}\frac{K}{4}\{\sin\alpha^e\sin\alpha^o\sin(\beta^e+\beta^o)\}+\frac{1}{4}\{h(\cos\alpha^e+\cos\alpha^o)+(\cos(\beta^e-\beta^o)+\gamma\cos(\beta^e+\beta^o))\sin\alpha^e\alpha^o\}
\end{equation}
\end{widetext} 
After performing minimization over the real parameters of \(\alpha^{e(o)}\) and \(\beta^{e(o)}\), the optimum value of \(\delta\) is obtained for
\begin{eqnarray}
    \nonumber\beta^e &=& \beta^o=\beta=\frac{1}{2}\arctan{\frac{K}{\gamma}},\\
    \nonumber \alpha^e &=& \arccos\big(\frac{1}{\omega}\big),\\
 \text{and} \quad   \alpha^o &=& \arctan \big (\sqrt{\omega^2-1}\big),
\end{eqnarray}
where \(\omega=(K\sin2\beta+\gamma\cos2\beta+1)/h\). It shows that \(\ket{\Phi}\) is the ground state of the Hamiltonian if \(\delta=\delta_g\) where \(\delta_g=-\frac{1}{2}\sqrt{h^2+K^2+\gamma^2}\) is the energy of ground state of \(H^{eo}\). This leads to the equation of factorization surface, which depends on the Hamiltonian parameters, 
\begin{equation}
    h^2=1-\gamma^2-K^2.
\end{equation}

\section{Ground state in non-Hermitian system}
\label{sec:ground_state_non_herm}

In our paper, we consider a Hamiltonian that exhibits two distinct regimes:  (i) an unbroken phase and  (ii) a broken phase. In the unbroken phase, the entire spectrum is real, and, therefore, the lowest-energy eigenstate can be identified unambiguously.  In contrast, in the broken phase, the eigenvalues become complex, and as a consequence, an additional criterion is required to determine the appropriate initial state for the dynamics. In that phase, we consider the imaginary eigenvalue with the lowest imaginary contribution to be the corresponding ground state in the broken phase. The time scale required to reach the steady state in the broken phase plays a crucial role, and is governed by the energy spectrum of the model.  The eigenvalues corresponding to each momentum sector $H_p^{\text{iKSEA}}$ are $\pm \epsilon_p$, where
\begin{equation}
\epsilon_p = \sqrt{(h+\cos\phi_p)^2-(\gamma^2-K^2)\sin^2\phi_p},
\end{equation}
and \(\phi_p=\frac{2p-1}{N}\) for $p=1,2,\dots, N/2$. The time evolution of the state in a given momentum sector \(p\) is described by
\begin{equation*}
\ket{\psi_p'(t)} = e^{-i H_p t}\ket{\psi_p(0)}
\end{equation*}
\begin{equation}
\ket{\psi_p(t)} = \frac{\ket{\psi_p'(t)}}{\|\ket{\psi_p'(t)}\|} = \frac{e^{-it\sum_{\alpha=\pm}E_p^\alpha \ket{R_\alpha}\bra{L_\alpha}}
\ket{\psi_p(0)}}{\|\ket{\psi_p'(t)}\|},
\label{eq:evol_state_1}
\end{equation}
where \(\ket{R_\alpha}\) and \(\bra{L_\alpha}\) denote the right and left eigenvectors of the non-Hermitian Hamiltonian, respectively and \(\ket{\psi_p(0)}\) is an arbitrary state.

In the long-time limit \(t\to\infty\), the relaxation towards the steady state is entirely governed by the imaginary parts of the complex eigenvalues. Consequently, the characteristic time scale required to reach the steady state for each momentum mode \(p\) is determined by
\begin{equation}
t \gg \frac{1}{\mathrm{Im}\,[\epsilon_p]}.
\end{equation}
Therefore, for a finite system, the steady state is reached in the broken phase in time $(\min_p\{\mathrm{Im}\,[\epsilon_p]\})^{-1}$, where the minimum is taken over discrete momentum modes $\phi_p$. Approaching the thermodynamic limit in the broken regime, the discrete values of momenta approach the critical momentum $\phi_{p_c}$, (where $\epsilon_{p_c}=0$), i.e., $\phi_p\to\phi_{p_c}$ with both $\epsilon_p^2\to 0^{+}$ and $\epsilon_p^2\to 0^{-}$. The lowest imaginary value $\min_p\{\mathrm{Im}\,[\epsilon_p]\}\to0$ for $\epsilon_p^2\to 0^{-}$ limit. Therefore, the long time limit is given by $t\to\infty$ in the thermodynamic limit.

\bibliography{ixyksea}
\end{document}